\documentclass[prc,twocolumn,showpacs]{revtex4}
\usepackage{epsf}   
\usepackage{amssymb}
\usepackage{dcolumn}
\begin{document}
\title{Dipole and quadrupole polarizabilities of the pion}
\author{L.V. Fil'kov}
\email[E-mail: ]{filkov@sci.lebedev.ru} 
\author{V.L. Kashevarov}
\affiliation{Lebedev Physical Institute, Leninsky Prospect 53,
Moscow119991, Russia}
\vskip 0.5cm 

\begin{abstract}
Data on pion polarizabilities obtained in different experiments
are reviewed. The values of the dipole and quadrupole polarizabilities of
the $\pi^0$ and $\pi^{\pm}$-mesons found are compared with predictions of
dispersion sum rules (DSRs) and two-loop calculations in the framework of 
chiral perturbation theory (ChPT).
Possible reasons of a difference between the predictions of DSRs and ChPT
are discussed.
\end{abstract}
\vskip 0.2cm

\pacs{{12.38.Qk} {Experimental test}, 
{13.40.-f} {Electromagnetic processes and properties}, 
{11.55.Fv} {Dispersion relations},  
{12.39.Fe} {Chiral Lagrangians}}

\maketitle
\vskip 0.5cm

By now the values of the pion polarizabilities were determined 
by analyzing the
processes $\pi^-A\to \gamma\pi^-A$, $\gamma p\to \gamma \pi^+ n$, 
and $\gamma\gamma\to\pi\pi$.

At present the most reliable method of the determination of the $\pi^0$
polarizabilities is an analysis of the process $\gamma\gamma\to\pi^0\pi^0$.
With this aim dispersion relations (DRs) at fixed $t$ with one subtraction
at $s=\mu^2$ (where t(s) is the square of total energy (momentum transfer) 
for the process under consideration, $\mu$ is the pion mass)
were constructed for the helicity amplitudes of this process \cite{fil1}.
Via the cross symmetry these DRs are identical to DRs with two subtractions.
The subtraction functions were determined with the help of DRs at fixed
$s=\mu^2$ with two subtractions and the subtraction constants were expressed 
through the sum and the difference of
the electric and magnetic dipole and quadrupole pion polarizabilities. 
It is worth to note that these DRs have not any expansions and so they
can be used for a determination of the polarizabilities in the region of 
both low and intermediate energies.

These DRs were used to fit the experimental data \cite{mars} to the total 
cross section of the
process $\gamma\gamma\to\pi^0\pi^0$ in the energy region 270--2250 MeV. 
The values of the dipole and quadrupole polarizabilities 
(in units $10^{-4}$fm$^3$ and $10^{-4}$fm$^5$, respectively) 
found in the fits \cite{fil1,kash} are listed in Table 1 
together with the results obtained in Ref. \cite{kal1,kal2} 
and the prediction of DSRs \cite{fil1} and two loop
calculations in the frame of ChPT \cite{bel,gas1}.
\begin{table*}
\caption{The dipole and quadrupole polarizabilities of the $\pi^0$ meson}
\centering
\begin{tabular}{|c|c|c|c|}\hline
        &fit\cite{fil1,kash} &DSRs \cite{fil1} &ChPT  \\ \hline
$    $   &$-1.6\pm 2.2$ \cite{fil3} &$-3.49\pm 2.13$ &$-1.9\pm 0.2$
\cite{bel}  \\ 
$(\alpha_1-\beta_1)_{\pi^0}$ &$-0.6\pm 1.8$ \cite{kal1}   &               
& \\ \hline
$(\alpha_1+\beta_1)_{\pi^0}$ &$0.98\pm 0.03$ \cite{kash} &
$0.802\pm 0.035$&$1.1\pm 0.3$ \cite{bel}  \\ 
        &$1.00\pm 0.05$ \cite{kal2} &                & \\ \hline
$(\alpha_2-\beta_2)_{\pi^0}$ &$39.70\pm 0.02$\cite{fil1}&
$39.72\pm 8.01$ &$37.6\pm 3.3$
\cite{gas1}       \\ \hline
$(\alpha_2+\beta_2)_{\pi^0}$ &$-0.181\pm 0.004$\cite{fil1}&
$-0.171\pm 0.067$& 0.04 \cite{gas1}       \\ \hline
\end{tabular}
\end{table*}
The obtained values of the dipole polarizabilities and the difference of the
quadrupole polarizabilities do not conflict within the errors with the 
predictions of DSRs and ChPT. 

As for the sum of the
quadrupole polarizabilities of $\pi^0$, the DSR calculation
agrees well with the experimental value, but ChPT predicts a positive
value in contrast to the experimental one. However, as it was noted in
Ref. \cite{gas1}, this quantity was obtained in  a two-loop approximation,
which is a leading  order result for this sum, and one expects substantial
corrections to it from three-loop calculations.

It should be noted that the values of the difference and the sum of the
quadrupole polarizabilities found from the fit have very small
errors. These are the fitting errors.
This is a result of the very high sensitivity of the total cross
section of the process $\gamma\gamma\to\pi^0\pi^0$
at $\sqrt{t}>1500$ MeV to these
parameters. In order to estimate the real values of the errors, model 
uncertainties should be added.

Recently,
an experiment on the radiative $\pi^+$ meson photoproduction from the proton
($\gamma p\to\gamma\pi^+ n$) was carried out at the Mainz Microtron MAMI in 
the kinematics region 537~MeV~$<E_{\gamma}<817$~MeV, 
$140^{\circ}\le\theta_{\gamma\gamma^{\prime}}\le 180^{\circ}$ with the aim 
to determine the dipole 
polarizabilities of the charged pion \cite{ahrens}. The difference of 
the electric and 
magnetic dipole polarizabilities of the $\pi^+$-meson have been determined
from a comparison of the data with the predictions of two different
theoretical models, the first one being based on an effective pole model
with pseudoscalar coupling while the second one is based on diagrams
describing both resonant ($\Delta(1232$, $P_{11}(1440)$, $D_{13}(1520)$,
$S_{11}(1535)$, and $\sigma$-meson) and nonresonant contributions. 
The validity of the models has been verified by comparing the predictions 
with the present experimental data in the kinematics region where the pion
polarizability contribution is negligible ($s_1<5\mu^2$, where $s_1$ is 
the square of the total energy in c.m.s. of the process 
$\gamma\pi\to\gamma\pi)$
and where the difference between the predictions of the two models
does not exceed 3\%. In the region, where the pion
polarizability contribution is substantial ($5<s_1/\mu^2<15$,
$-12<t/\mu^2<-2$, where $t$ is the square of the momentum transfer in 
c.m.s. of the process $\gamma p\to\gamma\pi^+ n$ ), 
the difference $(\alpha_1-\beta_1)_{\pi^+}$
of the electric and the magnetic dipole polarizabilities of $\pi^+$
has been determined: 
\begin{equation}
(\alpha_1-\beta_1)_{\pi^+} =11.6\pm 1.5_{stat}\pm 3.0_{syst}\pm 0.5_{mod}. 
\end{equation}
This result is in good agreement with the DSRs prediction \cite{fil1},
however, it is at variance with ChPT calculations \cite{gas2,burgi}.

An additional independent analysis \cite{igor} of the experimental data
\cite{ahrens}
was carried out by a constrained $\chi^2$ fit \cite{silin}.
A series of seven analyses with $\chi^2<4,\,5,\,6,\,7,\,8,\,9,\,10$
has been performed. The value for $(\alpha_1-\beta_1)_{\pi^+}$ stabilizes for
$\chi^2<5$. 
The result obtained
agrees very well with the first analysis giving it additional support.

The analyses of the reaction $\gamma\gamma\to\pi^+\pi^-$ with the
aim to determine the charged pion dipole polarizabilities have been performed
early in the energy region below 700 MeV \cite{kal1,bab,don}. 
However, in this region the values
of the experimental cross section of this process \cite{dm,mark} are very
ambiguous. As a result, the values of $(\alpha_1-\beta_1)_{\pi^\pm}$ found
lie in the interval 4.4--52.6. 
The analyses \cite{kal1,don} of the data of Mark II \cite{mark} only
have given $\alpha_{1\pi^{\pm}}$ close to the ChPT result. However,
even changes of this value by more than 100\% are still
compatible with the present error bars in the energy region considered
\cite{don}.

A new analysis of this process \cite{fil2} 
has been carried out in the energy region 280--2500 MeV using DRs with
subtractions similar to those used for the analysis of the process
$\gamma\gamma\to\pi^0\pi^0$. But in this case the Born amplitude is not 
equal to 0. These DRs, where the charged pion dipole and quadrupole 
polarizabilities 
were free parameters, were used to fit the experimental data for the total 
cross section \cite{mark,tpc} in the energy region under consideration. 
The values of the polarizabilities found in this work and the predictions of
DSRs and ChPT are listed in Table 2.
\begin{table*}
\caption{The dipole and quadrupole polarizabilities of the charged pions.}
\centering
\begin{tabular}{|c|l|l|c|c|} \hline
         &                       &                &\multicolumn{2}{|c|}{
ChPT \cite{gas2}} \\ \cline{4-5}
         &\qquad fit \cite{fil2} &DSRs \cite{fil1} &to one-loop & to two-loops
\\ \hline
$(\alpha_1-\beta_1)_{\pi^\pm}$   &$13.0^{+2.6}_{-1.9}$   
&$13.60\pm 2.15$ &6.0 &5.7 [5.5]  \\ \hline
$(\alpha_1+\beta_1)_{\pi^\pm}$   &$0.18^{+0.11}_{-0.02}$ 
&$0.166\pm 0.024$&0 & 0.16 [0.16] \\ \hline
$(\alpha_2-\beta_2)_{\pi^\pm}$   &$25.0^{+0.8}_{-0.3}$   
&$25.75\pm 7.03$ &11.9&16.2 [21.6] \\ \hline
$(\alpha_2+\beta_2)_{\pi^\pm}$   &$0.133\pm 0.015$       
&$0.121\pm 0.064$&0&-0.001 [-0.001]
\\ \hline
\end{tabular}
\end{table*}
The numbers in brackets correspond to another determination of low energy
constants (LECs) \cite{bijn} for the order $p^6$. 
As seen in this Table, all values of the polarizabilities found in Ref.
\cite{fil2} are in good agreement with the DSR predictions \cite{fil1}.
On the other hand, all these data, except for $(\alpha_1+\beta_1)_{\pi^\pm}$,
are at variance with the ChPT calculations. It should be noted that the LECs
at order $p^6$ are not well known and the two-loop  contribution to the
difference of the quadrupole polarizabilities is very big ($\sim 100\%$).
Therefore, the contribution of the three-loop calculations could be
considerable.

The first measurements of the pion dipole polarizabilities  have been
carried out by analyzing the radiative $\pi^-$-meson scattering off the
Coulomb field of heavy nuclei \cite{antip} ($\pi^-A\to \gamma\pi^-A$), 
which is similar to the well known Primakoff effect \cite{prim}.
It was shown \cite{faldt} that in this reaction the Coulomb amplitude
dominates for momentum transfer $|t|\lesssim 10^{-4}$ (GeV/c)$^2$.
In the region of 
$|t|\sim 10^{-3}$ (GeV/c)$^2$ Coulomb and nuclear contributions are of 
similar size. In this region the nuclear contribution, in particular
an interference between the Coulomb and nuclear amplitudes, should be taken
into account. At $|t|\sim 10^{-2}$ the nuclear contribution dominates.
In the Coulomb region the cross section of the process under review is usually
expressed by the Born cross section of the Compton scattering on the 
pion and an interference of the Born amplitude with the pion polarizabilities.
The relative contribution of the polarizabilities is maximum in scattering
of the photon at $\theta_{\gamma\gamma^{\prime}}\sim 180^{\circ}$,
in the $\gamma\pi$ c.m.s., and increases with the photon
energy. However, if we consider the process $\gamma\pi\to\gamma\pi$ even
below the $\rho$-meson production threshold, the corrections to the Born
amplitude could be essential. As shown in Ref. \cite{kash}, if the
total energy, in the $\gamma\pi$ c.m.s., $\sqrt{s}\gtrsim 450$ MeV 
and $\theta_{\gamma\gamma^{\prime}}\sim 180^{\circ}$, the $\sigma$-meson 
contribution is considerable.

In the work \cite{antip} the authors considered 
$|t|<6\times 10^{-4}$ (Gev/c)$^2$ and $\sqrt{s}<430$ MeV. Events in the region
of $|t|$ of $(2-8)\times 10^{-3}$ were used to estimate the nuclear background.
Assuming $(\alpha_1+\beta_1)_{\pi^-}=0$, the authors have obtained
\begin{equation}
(\alpha_1-\beta_1)_{\pi^-}=13.6\pm 2.8_{stat}\pm 2.4_{syst}. 
\end{equation}
This value is in good
agreement with the results of the works \cite{ahrens,fil2,fil3}.

In a more complete analysis of these experimental data \cite{antip2} these
authors have found 
\begin{equation}
(\alpha_1+\beta_1)_{\pi^-}=1.4\pm 3.1_{stat}\pm 2.5_{syst},
\end{equation}
\begin{equation}
(\alpha_1-\beta_1)_{\pi^-}=15.6\pm 8.7_{stat}\pm 6.1_{syst}.
\end{equation}

Let us consider possible reasons of the difference between the predictions of
DSRs and ChPT. The DSRs for the dipole polarizabilities were constructed 
using DRs
at fixed $u=\mu^2$ for the helicity amplitudes $M_{++}$ and 
$M_{+-}$ without subtractions (where $u$ is the square of the total energy in
c.m.s. in the cross channel of the Compton scattering on the pion).
In order to construct DSRs for the quadrupole polarizabilities, DRs with one
subtraction were used for the same amplitudes.
The main contribution to the DSRs for the differences of the electric
and magnetic polarizabilities of charged pions is given by the $\sigma$-meson.
However, this meson is taken into account in the present ChPT 
only partially through two-loop calculations.
Therefore, the predictions of DSRs and ChPT for this difference are so various.

In the case of the difference of the dipole polarizabilities of the 
$\pi^0$-meson, the big contribution of the $\sigma$-meson to DSRs
is cancelled by the big contribution of the $\omega$-meson. 
On the other hand, the $\sigma$-meson is only partially included
in the framework of ChPT, and the $\omega$-meson gives a very small 
contribution to this difference. As a result, the DSRs and ChPT predictions
for the difference of the dipole polarizabilities of the $\pi^0$-meson
are rather close.

Consider the methods of a calculation of the vector meson contribution 
in the frameworks of DSR and ChPT. The Breit-Wigner expression for the vector meson 
contribution to the helicity amplitude $M_{++}(s,t)$ can be written as
\begin{equation}
M_{++}(s,t)=\frac{-4g_{\gamma\pi}^2s}{(m_v^2-s-i\Gamma m_v)}.
\end{equation}
In the narrow width approximation we have 
\begin{equation}
Im M_{++}(s,t)=
-4\pi g_{\gamma\pi}^2s \delta (s-m_v^2).
\end{equation}
Then in the framework of DSR we obtain
\begin{equation}\label{dsr}
Re M_{++}(s=\mu^2,t=0)=\frac{-4g_{\gamma\pi}^2m_v^2}{(m_v^2-\mu^2)}.
\end{equation}
This non-Born amplitude is expressed through the difference of the electric and
the magnetic dipole polarizabilities of the pion as 
\begin{equation}
Re M_{++}(s=\mu^2,t=0)=2\pi\mu(\alpha_1-\beta_1).
\end{equation}

In the case of ChPT the authors of Ref. \cite{bel} used
\begin{equation}\label{chpt}
Re M_{++}(s=\mu^2,t=0)=\frac{-4g_{\gamma\pi}^2\mu^2}{(m_v^2-\mu^2)}.
\end{equation}

The value of the amplitude (\ref{chpt}) is smaller than (\ref{dsr}) by
$m_v^2/\mu^2$ times.
From the point of view of analyticity, the result (\ref{chpt}) 
can be obtained
if a DR with one subtraction at $s=0$ is used for the amplitude $M_{++}(s,t)$.
However, an additional subtraction constant appears then, which was
not considered in the available ChPT calculations. This leads to an additional
disagreement between the predictions of DSRs and ChPT for 
$(\alpha_1-\beta_1)_{\pi^\pm}$. 
\vspace{0.1cm}

The authors thank D. Drechsel, J. Gasser, M.A. Ivanov, and Th. Walcher  
for useful discussions.
This research is part of the EU integrated initiative hadron physics project
under contract number RII3-CT-2004-506078 and was supported in part by the
Russian Foundation for Basic Research (Grant No. 05-02-04014).

%


\begin{thebibliography}{99}
\bibitem{fil1} L.V. Fil'kov and V.L. Kashevarov, Phys. Rev. C {\bf 72}, 035211
(2005).
\bibitem{mars} H. Marsiske {\em et al.}, Phys. Rev. D {\bf 41}, 3324 (1990);
J.K. Bienlein, Crystal Ball Contribution to the 9th Intern.
Workshop on Photon-Photon Collisions, San Diego, California, 22-26 March 1992.
Proceedings: Photon-Photon Collisions, edited by D.O. Caldwell and
H.P. Paar, River Edge, N.Y., Word Scientific, 1992, p.241.
\bibitem{kash} L.V. Fil'kov and V.L. Kashevarov, Eur. Phys. J. A {\bf 5},
285 (1999).
\bibitem{kal1} A.E. Kaloshin and V.V. Serebryakov, Z. Phys. C {\bf 64}, 689
(1994).
\bibitem{kal2} A.E. Kaloshin, V.M. Persikov, and V.V. Serebryakov, Phys. Atom.
 Nucl. {\bf 57}, 2207 (1994).
\bibitem{bel} S. Bellucci, J. Gasser, and M.E. Sainio, Nucl. Phys. B {\bf 423},
80 (1994); B {\bf 431}, 413 (1994).
\bibitem{gas1} J. Gasser, M.A. Ivanov, and M.E. Sainio, Nucl. Phys. B {\bf 728},
31 (2005).
\bibitem{ahrens} J. Ahrens {\em et al.}, Eur. Phys. J. A {\bf 23}, 113 (2005).
\bibitem{gas2} J. Gasser, M.A. Ivanov, and N.E. Sainio,
Nucl. Phys. B {\bf 745}, 84 (2006).
\bibitem{burgi} U. B\"urgi, Nucl. Phys. B {\bf 479}, 392 (1997).
\bibitem{igor} I.~Giller, Ph.D. thesis, Tel Aviv University, (2004).
\bibitem{silin} S.N. Dymov, V.S. Kurbatov, I.N. Silin, and S.V. Yaschenko,
Nucl. Inst. Meth. {\bf A440}, 431 (2000).
\bibitem{bab} D. Babusci {\em et al}., Phys. Lett. B {\bf 277}, 158 (1992);
\bibitem{don} J.F. Donoghue and B.R. Holstein, Phys. Rev. D {\bf 48}, 137 (1993).
\bibitem{fil2} L.V. Fil'kov and V.L. Kashevarov, Phys. Rev. C {\bf 73}, 035210
(2006).
\bibitem{dm} PLUTO Collaboration (C. Berger {\em et al}.), Z. Phys. C
{\bf 26} 199 (1984);
DM1 Collaboration (A. Courau {\em et al}.), Nucl. Phys. B {\bf 271}, 1 (1986);
DM2 Collaboration (Z. Ajaltoni {\em et al}.),
Phys. Lett. {\bf B194}, 573 (1987).
\bibitem{mark} Mark II Collaboration (J. Boyer {\em et al}.),Phys. Rev. D
{\bf 42}, 1350 (1990).
\bibitem{tpc} TPC/$2\gamma$ Collaboration (H. Aihara {\em et al.}),
Phys. Rev. Lett. {\bf 57}, 404 (1986);
CELLO Collaboration (H.J. Behrend {\em et al.}), Z. Phys.
C {\bf 56}, 381 (1992);
VENUS Collaboration (Fumiaki Yabuki {\em et al.}),
J. Phys. Soc. Jap. {\bf 64}, 435 (1995);
ALEPH Collaboration (A. Heister {\em et al.}), Phys. Lett.
B {\bf 569}, 140 (2003);
Belle Collaboration (H. Makazawa {\em et al.}), Phys. Lett.
B {\bf 615}, 39 (2005).
\bibitem{bijn} J. Bijnens and J. Prades, Nucl. Phys. B {\bf 490}, 239 (1997).
\bibitem{fil3} T.A. Aibergenov {\em et al.}, Czech. J. Phys. B {\bf 36}, 948 
(1986).
\bibitem{antip} Yu.M. Antipov {\em et al.}, Phys. Lett. B {\bf 121}, 445 (1983).
\bibitem{prim} H. Primakoff, Phys.Rev. {\bf 81}, 899 (1951);
I.Ya. Pomeranchuk and I.M. Shmushkevich, Nucl. Phys. {\bf 23}, 452 (1961).
\bibitem{faldt} G. F\"{a}ldt, Phys. Rev. C {\bf 76}, 014608 (2007).  
\bibitem{antip2} Yu.M. Antipov {\em et al.}, Z. Phys C {\bf 26}, 495 (1985).

\end{thebibliography}
\end{document}